\documentclass[preprint2]{aastex}

\shorttitle{The interacting galaxy pair KPG 390}
\shortauthors{Repetto et al.}

\begin{document}

\title{The interacting galaxy pair KPG 390: 
H$\alpha$ kinematics.}

\author{P. Repetto\altaffilmark{1}, M. Rosado\altaffilmark{1} 
R. Gabbasov\altaffilmark{1} and I. Fuentes-Carrera\altaffilmark{2}}
\affil{Instituto de Astronom\'{\i}a, Universidad Nacional 
Autonoma de M\'exico (UNAM),\\ Apdo. Postal 70-264, 04510, 
M\'exico, D.F., M\'exico}\affil{Department of Physics, Escuela 
Superior de F{\'i}sica y Matem\'aticas, IPN, U.P. Adolfo L\'opez Mateos, 
C.P. 07738, Mexico city IPN, M\'exico.}

\altaffiltext{1}{Send offprint requests to: P. Repetto} 
\email{prepetto@astroscu.unam.mx}

\begin{abstract}
In this work we present scanning Fabry-Perot
H$\alpha$ observations of the isolated interacting
galaxy pair NGC 5278/79 obtained with the PUMA Fabry-Perot interferometer.
We derived velocity fields and rotation curves for both galaxies.
For NGC 5278 we also obtained the residual velocity map to investigate
the non-circular motions, and estimated its mass by fitting the rotation
curve with a disk+halo components.
We test three different types of halo (pseudo-isothermal, Hernquist
and Navarro Frenk White) and obtain satisfactory fits to 
the rotation curve for all profiles.
The amount of dark matter required by pseudo-isothermal profile 
is about ten times smaller than, that for the other two halo distributions.
Finally, our kinematical results together with the analysis of dust lanes
distribution and of surface brightness profiles along the minor axis 
allowed us to determine univocally that both components of the 
interacting pair are trailing spirals.
\end{abstract}


\section{Introduction}

Interactions and mergers of galaxies are common
phenomena in the Universe. Isolated pairs of galaxies
represent a relatively easy way to study
interactions between galaxies because these systems,
from a kinematical point of view, are simpler than
associations and compact groups of galaxies, where
so many galaxies participate in the interaction process
that it is difficult to discriminate the role of each
galaxy in the interaction.

In the case of isolated disk galaxies the physical processes
that determine secular evolution are internal
processes (internal secular evolution).
The timescale of these processes
is of the order of many galaxy rotation periods
and generally involve the interactions of individual
stars or gas clouds with collective phenomena such as bars,
oval distorsions, spiral structure and triaxial
dark matter halos \citep{Kormendy2004, Kormendy2008}.
In the case of interacting galaxies one
talks about environmental secular evolution.
The time scale of the environmental secular
evolution is of the same order as that of the internal
secular evolution (both are slow processes).
This phenomenon is driven by prolonged
gas infall, by minor mergers and by galaxy
harassment.
Secular evolution processes (internal or external)
are slow phenomena in comparison with galaxy
mergers and stripping of gas that also belong
to the context of interacting galaxies and
determine their evolutionary stage \citep{Kormendy2004, Kormendy2008}.

Long-slit spectroscopy studies \citep{Nelson1998, Liu1995} 
restrict the information only to a few points in a galaxy.
These kinds of studies are principally focused on
axisymmetric systems such as isolated galaxies because a full
rotation curve is obtained by placing the slit along the major axis
of the galaxy. In the case of interacting systems, the slit is usually placed
only in a few positions obtaining spectra on regions of interest.
Although some kinematic information is derived, it is not possible
to obtain true rotation curves of the galaxies.
On the other hand, scanning
interferometric Fabry-Perot observations give us
kinematic information of the whole interacting system.
This is very important in the case of asymmetrical and
perturbed systems as is the case with interacting
galaxy pairs.
Extended kinematic information can help determine
the effect of the interaction process on each of
the members of the system
\citep{ Amram2004,Rampazzo2005,Fuentes-Carrera2007,Bizyaev2007}.
This technique allows more sensitivity to
detect faint objects due to increased brightness and field
of view and has more spatial and
spectral resolution than classic spectrographs.

Obtaining kinematical
information on interacting galaxies
systems is useful to understand
the effect the interaction could
have on each of the members of the pair
\citep{Marcelin1987, Amram1994, Amram2002a,
Fuentes-Carrera2004, Fuentes-Carrera2007}.
Rotation curves measurement, for instance,
is necessary to study the
mass distribution in spiral galaxies and for estimation of the
amount of dark matter
\citep[][among other authors]{Rubin1976, Bosma1977, Blais$-$Ouellette2001}.

There are at least two independent methods to
study the mass distribution in paired galaxies
using kinematics. The first one considers the difference
in systemic velocities of the galaxy components as
a lower limit to the orbital velocity of the smaller
galaxy (companion) assuming a circular orbit,
and the projected separation of the orbital radius of the
companion galaxy, in order to estimate the lower limit mass
of the larger (primary) galaxy \citep{Karachentsev1984}.
The second one is
based on the decomposition of the rotation curve,
considering the contribution of various mass components such as bulge,
disk and dark matter halo \citep{van$-$Albada1985}.
Thus, one can compare the mass derived from
the orbital velocity of the companion in a galaxy
pair to the more detailed predictions of a mass model,
based on the kinematics of each galaxy component.

In this work we study the system NGC 5278/79 (known also as Arp 239 and KPG 390) belonging
to a particular class of interacting pair of galaxies: the M51-type
galaxies. According to \citet{Reshetnikov2003}, the two
empirical criteria to classify an M51-type pair of galaxies are that the
B-band luminosity ratio of the components (main/satellite) vary between $1/30$
and $1/3$, and that the projected distance of the satellite
does not exceed two optical diameters of the main component.
In the case of NGC 5278/79 the B-band luminosity ratio is $0.30$ and the
projected separation is $16.8$ kpc (optical diameter
of the main component is $39.2$ kpc as we can see from
Table \ref{tbl-1}). Thus, NGC 5278/79 is classified as an
M51-type pair of galaxies. M51-type galaxy pairs are interesting
because the mass of the system can be evaluated, in principle, by two different ways,
mentioned above: by means of the rotation curves of each
component, and by estimating the orbital motion of the satellite galaxy
around the main galaxy. 
With the help of HST images \citep{Windhorst2002},
showing dust lanes across
the nuclei of both galaxy components we can determine
which are the nearest sides of the galaxies and thus, to determine
whether the spiral arms are leading or trailing, as well as to have
an extensive view of the geometric conditions of the encounter in
order to perform numerical simulations. It is important to recall
that according to several studies, leading spiral arms
can only exist in interacting systems with retrograde encounters
triggering the formation of $m=1$ spiral arms \citep{Athanassoula1978,
Byrd1989, Thomasson1989, Keel1991, Byrd1993}, conditions
that, in principle, taking into account the morphology,  could be
fulfilled by  NGC 5278/79. Thus, it is worth to explore this possibility
in this particular case. These are the main motivations of this work.

In this paper we present
scanning Fabry-Perot observations, velocity fields
and rotation curves of this interacting galaxy
pair. The aim of this study is to perform detailed kinematic
and dynamic analysis of NGC 5278/79
using H$\alpha$ kinematical data in order to study
the mass distribution of this pair of galaxies with the
two methods mentioned above and to determine the type of spiral arms
(leading or trailing) in
the galaxy members with the intention of reproducing both its morphology and
kinematics with future numerical simulations that could  shed more light
on the interaction process.
Last, but not least, there are no
3D spectroscopic works on this pair (neither scanning H$\alpha$
Fabry-Perot nor HI) neither X-ray data. 
In Section \ref{datan} we
present an overview of the observations
and the data reduction process.
In Section \ref{resl} we present the derived velocity field and
the associated rotation curve of each component of the pair.
In Section \ref{orm} the kinematics inferred from the velocity
fields and rotation curves is discussed analyzing
carefully the role of non-circular motions.
Section \ref{mcrcd} is devoted to the dynamical analysis
with the computation of the mass for each galaxy
of the pair. The discussion is presented in Section \ref{dsc},
and conclusions are presented in Section \ref{cls}.

\subsection{NGC 5278/79 (Arp 239; KPG 390; Mrk 271)}

This pair was
first cataloged by \citet{Vorontsov-Velyaminov1959}
with the identifier VV 019. It was later included in Arp's
{\it Atlas of Peculiar Galaxies} with the identifier
Arp 239 \citep{Arp1966}. Grouped with other objects
it was classified as {\it appearance of fission}.
It appears in the Karachentsev's catalog of isolated
pair of galaxies \citep{Karachentsev1972} as KPG 390.
According to Karachentsev's classification, NGC 5278 is the
primary galaxy of the pair (i.e., the main component) and
NGC 5279 is the secondary one (i.e., the satellite, according
to the discussion on M51-type of galaxy pairs given above).
In the {\it Uppsala General Catalog of Galaxies} NGC 5278
and NGC 5279 are identified as UGC 08677 and UGC 08678,
respectively \citep{Nilson1973}.
The main data about this pair are collected in
Table \ref{tbl-1}. There are not many works on
this pair in the literature. The existing works are of
statistical character mostly, as part of vast surveys
of interacting galaxies \citep{Turner1976, Cutri1985,
Klimanov2001}.

\begin{figure}[!htp]
\plotone{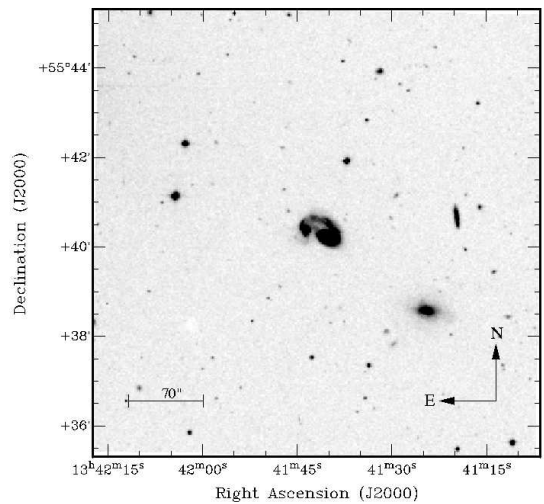} \figcaption{DSS blue band image of Arp 239.
The north-east direction
and the scale are indicated in the figure. \label{fig1}}
\end{figure}

\begin{figure}[!htp]
\plotone{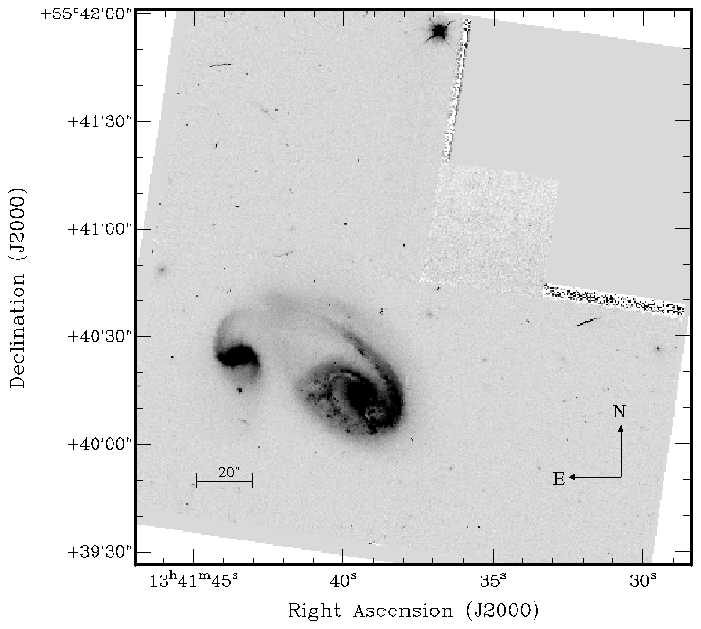} \figcaption{HST image at $\lambda=
8230$ \AA\, with the F814W filter \citep{Windhorst2002}. \label{fig2}}
\end{figure}

The pair consists of two spiral galaxies NGC 5278 and NGC 5279
(Fig.~\ref{fig1} and Fig.~\ref{fig2}). In the DSS blue band image shown 
in Fig.~\ref{fig1} is evident the bridge region, fainter 
than the continuum emission of the two components of KPG 390.

Figure \ref{fig2} shows HST image from \citet{Windhorst2002}.
These authors, by means of images observed with the HST Wide Field
and Planetary Camera 2 (WFPC2) at several UV-wavelengths, notice that the inner parts of both components show
significant dust lanes and observe that part of
the dust seems to spread along one of the arms. The most remarkable
feature is a very curved thin dust
lane that drapes across the primary galaxy. This dust
lane starts near the southern spiral arm of NGC 5278,
curves around the small nuclear bulge of this galaxy,
and appears in the spiral arm connecting the two galaxies
in the north-east direction as one can see in Fig.~\ref{fig2}.
The presence of dust lanes in this M51-type pair
allows to determine the tilt of the galaxy (nearer or farther
side) once we know the kinematics as it will be discussed later on.

The HST images also show that
the encounter clearly distorts the galaxy
disks of both components. The result is the formation of
tidal tails and excitation of a strong $m=1$ mode
in NCG 5278.

\begin{table*}[!htp]
\begin{minipage}[t]{\columnwidth}
\caption{Parameters of NGC 5278 and NGC 5279.}
\label{tbl-1}
\begin{center}
\renewcommand{\footnoterule}{}  
\begin{tabular}{lcr}
\tableline\tableline
Parameters & NGC 5278 & NGC 5279\\
\tableline\tableline
Coordinates (J2000)\tablenotemark{a}
& R.A.$=13^{h}41^{m}39.6^{s}$
& R.A.$=13^{h}41^{m}43.7^{s}$ \\
& Dec.$=+55\degr 40\arcmin 14\arcsec$
& Dec.$=+55\degr 40\arcmin 26\arcsec$ \\
Morphological type \tablenotemark{b}  & SA(s)b? pec  & SB(s)a pec \\
Distance (Mpc) \tablenotemark{c}  & 100.6  & 101.1 \\
Apparent diameter (arcsec)\tablenotemark{c} & 80.9$\pm$12.0 & 36.20$\pm$6.30 \\
Apparent diameter (kpc) \tablenotemark{c} & 39.2$\pm$6.0 & 17.5$\pm$3.10 \\
Axis ratio\tablenotemark{c}& 0.71  & 0.72 \\
$m_{B}$ (mag)\tablenotemark{e}  &  14.29   & 15.20 \\
$M_{B}$ (mag)\tablenotemark{e} & -20.95  & -20.05 \\
Surface brightness (mag arcsec$^{-2}$) \tablenotemark{e} & 22.51  & 22.96 \\
Systemic velocity (km s$^{-1}$) & 7541\tablenotemark{c} 
& 7580\tablenotemark{c} \\
& 7627\tablenotemark{d} & 7570\tablenotemark{d} \\
Major axis position angle (deg)& 53.0$\degr$\tablenotemark{f} 
& 3.0$\degr$\tablenotemark{f} \\
& 42.0$\degr$\tablenotemark{d} & 141.5$\degr$\tablenotemark{d} \\
Inclination (deg)& 38.9$\degr$\tablenotemark{f} 
& 59.8$\degr$\tablenotemark{f} \\
& 42.0$\degr$\tablenotemark{d}
& 39.$\degr$6\tablenotemark{d} \\
\tableline\tableline
\end{tabular}
\tablenotetext{a}{\citet{Adelman-McCarthy2008, Jarrett2003}}.
\tablenotetext{b}{\citet{vandenBergh2003}}.
\tablenotetext{c}{\citet{de$-$Vaucouleurs1991}}.
\tablenotetext{d}{This work}.
\tablenotetext{e}{\citet{Mazzarella1993}}.
\tablenotetext{f}{\citet{Paturel2000}}.
\end{center}
\end{minipage}
\end{table*}

\citet{Keel1985} present spectra of NGC 5278/9, in an attempt
to investigate the nuclear activity of the pair, and
classify this interacting pair as an intermediate
class between Seyfert 2 nuclei and LINER (both galaxies).
Radio observations in 21cm HI line found a HI flux of $1.71$ Jy
km s$^{-1}$ and a total HI mass of $2.8\times10^{10}$ M$_\odot$ \citep{Bushouse1987}.
Recently, \citet{Condon2002} identified KPG 390 as a radio source stronger than
$2.5$ mJy at $1.4$ GHz from the NRAO VLA sky Survey (NVSS). These authors
give a $1.4$ GHz flux density of $23.1$ mJy.
The broad-band CCD images of this pair presented by \citet{Mazzarella1993}
reveal a number of regions
of enhanced brightness, being the brightest one located in NGC 5279.
These authors find morphological properties, luminosities and colors
of this system, and also give photometry of the individual
nuclei and giant HII regions of KPG 390 or Mrk 271.
As this pair is labeled as Markarian, it is 
very prominent in GALEX images \citep{Martin2005}. 
A glimpse at FUV and NUV images
reveals a bridge between the two galaxies, being much brighter in NUV.
Such UV emission is generally associated with regions of 
intense star formation. Thus, the star formation along the bridge 
may be induced by a recent tidal passage. In the disk of NGC 5278
and NGC 5279 the emission is irregular and no apparent
gradient of emission is observed.

\citet{Mazzarella1993} use long-slit
spectrometers of low spectral resolution
to derive spectroscopic data of KPG 390. 
\citet{Klimanov2002} by means of long-slit spectroscopy find
that both galaxies rotate in opposite directions.
An interesting detail noted by those authors is a large
discrepancy between dynamic and photometric centers
of both components (off centered in positions:
$3\arcsec$  ($1.4$ kpc) for NGC 5278, and $6\arcsec$ ($2.9$ kpc)
for NGC 5279).

Concerning the closest environment of the pair,
in the west side of Fig.~\ref{fig1} (field of $10\arcmin$) 
are clearly visible two
neighbour galaxies: UGC 8671 seen as edge-on elliptical 
and MCG+09-22-094.
The radial velocity of UGC 8671 is $7589$ km s$^{-1}$
\citep{Falco1999} and the radial velocity of
MCG+09-22-094 is $11700$ km s$^{-1}$ \citep{vandenBergh2003}.
According to \citet{Karachentsev1972}, the difference in
radial velocity for galaxies to be an interacting system
should be $\Delta V\leq500$ km s$^{-1}$.
For this reason MCG+09-22-094 is a false
neighbour and in the case of UGC 8671 $\Delta V\approx40$
km s$^{-1}$, so perhaps this galaxy is a member
of KPG 390. Althought the lack of information
on mass of UGC 8671, using the values of the major axis $33\arcsec$ 
\citep{Nilson1973}
and the B-band magnitude $14.50$ \citep{de$-$Vaucouleurs1991}  we
infer that the mass of UGC 8671 is very similar to the mass of NGC 5279
(see Table \ref{tbl-1}).
In order to determine if UGC 8671 is part of a group with
NGC 5278 and NGC 5279 we check whether this galaxy satisfies the
basic isolation criterion \citep{Karachentsev1972}. Karachentsev
examines galaxies with apparent magnitudes of pair members $m \leq 15.7$.
The parameters of the basic
isolation criterion are $\chi=5$, $\xi=0.5$, $\lambda=4$ and
the angular diameters of NGC 5278, NGC 5279, UGC 8671 are
a$_1=81\arcsec$, a$_2=36\arcsec$ and a$_3=33\arcsec$, respectively
(see Table \ref{tbl-1}). Applying the basic isolation criterion
to these data we see that the angular diameter of UGC 8671 does not
occur in the interval $\xi$ a$_1$ $\leq$ a$_3$ $\leq$ $\lambda$ a$_1$, 
it occurs instead in the interval $\xi$ a$_2$ $\leq$ a$_3$ $\leq$ 
$\lambda$ a$_2$. In the first case KPG 390 is an 
isolated pair of galaxies. In the second case it
satisfies the inequality $x_{23}/x_{12} \geq \chi $ a$_3$/a$_2$ and, 
consequently, this pair can also be considered as isolated according 
to the basic criterium. Thus, according to Karachentsev criteria,
the UGC 8671 does not belong to KPG 390. On the other hand there are
at least other thirtheen galaxies that are visible in larger field of 
$25\arcmin$ DSS blue image (not presented in this work).  
In the south-west direction there are two galaxies PGC 2507704
and PGC 2509387 respectively with radial velocities 10992 km s$^{-1}$
and 7358 km s$^{-1}$. 
In the north-east direction are located
other five galaxies PGC 2513261, PGC 2514229, 
SDSS J134224.97+554926.0, PGC 2515693, PGC 2516823, respectively with
radial velocities 7575 km s$^{-1}$, 20848 km s$^{-1}$, 12504 km s$^{-1}$
, 21573 km s$^{-1}$ and 20293 km s$^{-1}$. 
Other 
six galaxies are clearly visible in the south-east direction IC0922, 
PGC 2507810, PGC 2505734, IC 0918, IC 0919 and PGC 2505000, respectively 
with radial velocities 19955 km s$^{-1}$, 21006 km s$^{-1}$, 21328
km s$^{-1}$, 21183 km s$^{-1}$, 10555 km s$^{-1}$ and 10453 km s$^{-1}$
\citep{Abazajian2004}. The present identification is partial because 
there are other fainter and smaller galaxies that 
we cannot distinguish in the DSS blue image. 
The galaxies PGC 2509387 and PGC 2513261 have 
$\Delta V\leq200$ km s$^{-1}$ with respect to KPG 390, but
diameters $11.85\arcsec$ and $9.69\arcsec$ such that 
they do not satisfy the basic criterium of \citet{Karachentsev1972}. 

It would be useful to complement our
H$\alpha$ kinematic study with
EVLA HI observations with high spatial extent and resolution. 
The 21 cm radio observations 
could reveal extended features of ongoing interaction, providing 
additional information on kinematics of the KPG 390 environment. 

\section{Observations and data reduction}\label{datan}
Observations of NGC 5278/79 (Arp 239, KPG 390) were done in
2002 July at the f$/7.5$ Cassegrain focus of the 2.1 m
telescope at the Observatorio Astron\'omico Nacional
in San Pedro M\'artir (M\'exico), using the scanning
Fabry-Perot interferometer PUMA \citep{Rosado1995}.
We used a $1024\times1024$ Site CCD detector
and considered only the central $700\times700$
pixels corresponding to field of view of
$3.5\arcmin\times3.5\arcmin$
arcminutes, encompassing the two galaxy components. The resulting
image layout after
a $2\times2$ binning in both spatial dimensions
was $350\times350$ pixels with a spatial
sampling of $1.16\arcsec$ pixel$^{-1}$.
In order to isolate the redshifted H$\alpha$
emission, we used an interference filter centered
at $6750$ \AA \,with a FWHM of $50$ \AA.
PUMA is a focal reducer built at the Instituto de
Astronom\'{\i}a-UNAM used to make direct imagery
and Fabry-Perot (FP) spectroscopy of extended emission
sources (field of view $10\arcmin$). The FP used is an
ET-50 (Queensgate Instruments) with a servostabilization
system having a free spectral range of
$19.95$ \AA\, ($912$ km s$^{-1}$) at H$\alpha$.
Its finesse ($\sim24$) leads to a sampling spectral
resolution of $0.41$ \AA\, ($19.0$ km s$^{-1}$) which is
achieved by scanning the interferometer free spectral
range through 48 different channels \citep{Rosado1995}.

To average the sky variations during the exposure,
we got two data cubes with an exposure time of
$96$ min each ($120$ s per channel).
These data cubes were co-added leading to a total
exposure time of $192$ min. For calibration
we used a Ne lamp whose $\lambda=6717.043$ \AA\, was
close to the redshifted nebular wavelength.
Two calibration cubes were obtained at the beginning
and at the end to check the metrology.

The parabolic phase map was computed from the
calibration rings in order to obtain the reference
wavelength for the line profile observed inside
each pixel.
The instrumental and observational parameters are shown
in Table \ref{tbl-2}.

\begin{table}[ht]
\begin{minipage}[t]{\columnwidth}
\caption{Instrumental and observational parameters.}
\label{tbl-2}
\begin{center}
\renewcommand{\footnoterule}{}  
\begin{tabular}{l l}
\tableline
\tableline
Parameter & Value\\
\tableline
\tableline
Telescope   &	2.1 m (SPM)\\

Instrument   &	PUMA\\

Detector  &  Site3 CCD\\

Detector size (pixels) &  (1024$\times$1024)\\

Image scale ($2\times2$ binning)	&  1.16\,$''$\\

Scanning F-P interferometer	&  ET-50\\

F-P interference order at H$\alpha$	&  330\\

Free spectral range at H$\alpha$ (\AA)	&  19.95 \\

Spectral sampling at H$\alpha$ (\AA)  &  0.41 \\
Interference filter (\AA)   & 6750\\

Total exposure time (min) & 192\\

Calibration line (\AA) &  6717.043 (Ne)\\
\tableline
\tableline
\end{tabular}
\end{center}
\end{minipage}
\end{table}

The data reduction and most of the analysis were done using
the ADHOCw\footnote{http://www.oamp.fr/
adhoc/ developped by J. Boulesteix.}
software. Standard corrections were done on
each of the cubes: removal of cosmic rays
and bias subtraction. In this case the
object cubes do not show a significant
night sky continuum or OH sky lines pattern.
No spatial nor spectral smoothing was applied to
the data.
Through the scanning process, we obtained
for each pixel a flux value at each of the
48 velocity channels. The velocity profile
found along the scanning process contains
information about the monochromatic emission
(H$\alpha$) and the continuum emission of
the object. The continuum image computation
was done considering the mean of $3$ channels 
with lowest intensities.
For the H$\alpha$  image, the H$\alpha$
line intensity was obtained by adding those
channels (from 17 to 30) of the wavelength data cube for
which the diffuse bridge emission was stronger.
The seeing of the H$\alpha$ image is $2.3\arcsec$
and the signal to noise ratio $4$.
The velocity maps were computed using the barycenter
of the H$\alpha$ velocity profile at each pixel.
We mask the velocity field excluding a rectangular
area which contains the part of the radial velocity
map corresponding to KPG 390. Then we superimpose on
the resulting velocity field the radial velocity profiles
and delete every pixel that shows too low signal
to noise ratio. In addition, we build the FWHM map by
fitting Gaussians to every pixels of the wavelength
data cube.

\section{Results}\label{resl}
In this section we present the analysis of H$\alpha$ 
images and velocity fields of the KPG 390, and describe 
the obtained rotation curves.
\subsection{H$\alpha$ image}
\begin{figure}[!htp]
\epsscale{1.0}
\plottwo{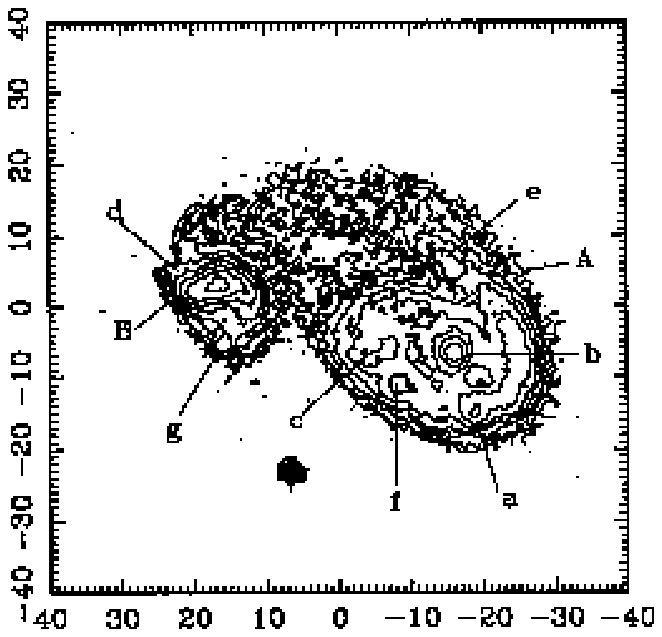}{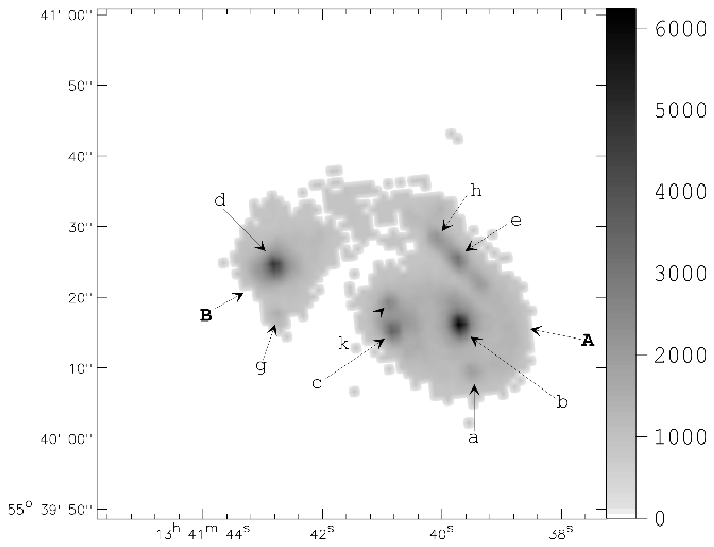} \figcaption{{\it Left},
HII regions identification by
\citet{Mazzarella1993}. Contour levels range from 26 to 17
mag arcsec$^{-2}$. {\it Right}, H${\alpha}$
image of NGC 5278/79, obtained in this work,
with identification of visible HII regions.
Capital letters refer to the components
of KPG 390.\label{fig3}}
\end{figure}

Figure \ref{fig3} shows the H$\alpha$
image of the pair KPG 390 obtained from
our FP interferograms, as described in
Sect.~\ref{datan}. We compare this map
with the image by \citet{Mazzarella1993}
(their Fig.1)\footnote{Reproduced by permission of the AAS.}.
Those authors present contour diagrams
of B band images and find at least five giant
HII regions, named with
small case letters {\it a}, {\it c}, {\it e},
{\it f} and {\it g}. The letters
{\it d} and {\it b} refer to the nuclei of the
two components.
In our monochromatic map at least
six giant HII regions are clearly visible, and
four of these regions coincide with zones
{\it g}, {\it c}, {\it a} and {\it e} of
\citet{Mazzarella1993}.
The giant HII region {\it f} is not visible in
our map, but there are other two giant HII
regions that are missing in the map of those authors
(k and h respectively in Fig.~\ref{fig3}).
In the presented monochromatic map the nuclei of the
two components are clearly visible,
named with the same convention of \citet{Mazzarella1993}.
Moreover, the region between the two components of the pair
is also present in our image.
This region is also visible in the B band
contour of \citet{Mazzarella1993}.
In the right panel of Fig.~\ref{fig3}, in the
northeastern side of the primary component of
the pair, we see clearly the bridge component
between the two galaxies probably
induced by the interaction. The bridge
begins at the end of the spiral arm of NGC 5278
where the brighter HII regions in the arm of NGC 5278
totally disappear and the H$\alpha$ emission
changes into fainter diffuse emission. The bridge region
ends in the proximity of the north-western part of the
disk of NGC 5279.

\subsection{Velocity fields}\label{velf}
\begin{figure*}[!htp]
\epsscale{2.0}
\plottwo{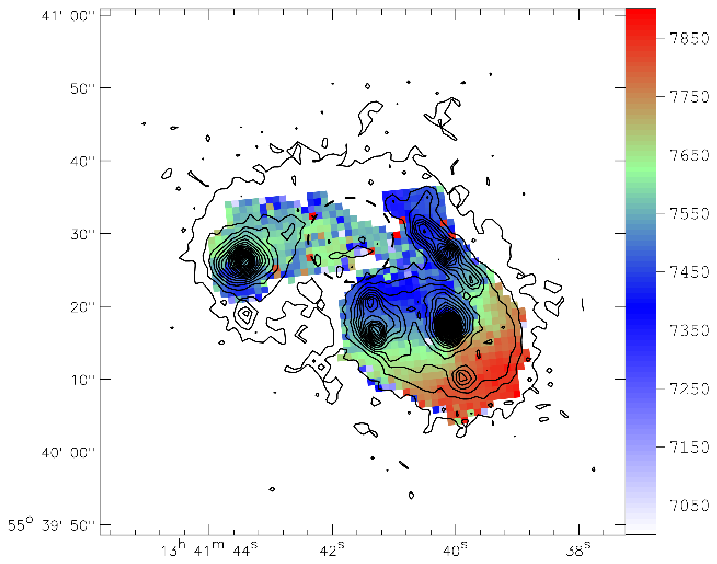}{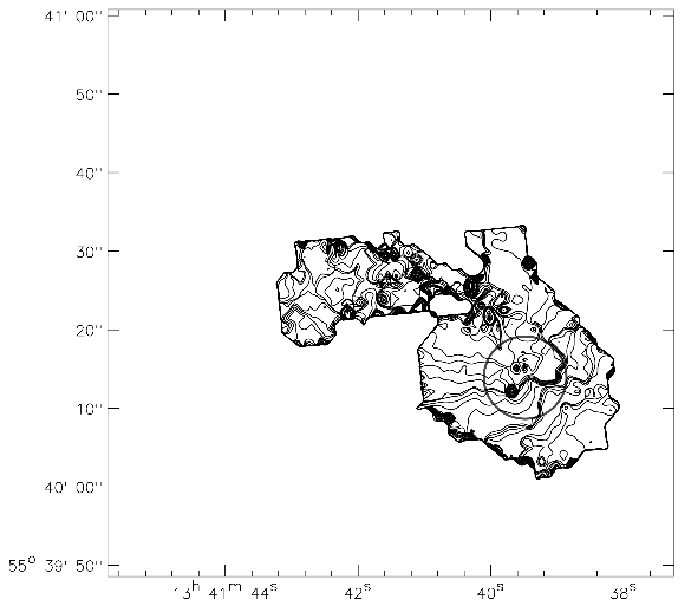} \figcaption{
{\it Left}: The full velocity field of KPG 390 (Arp 239)
with overplotted H$\alpha$ image isophotes. The isophotes 
are separated by a factor of $200$ in arbitrary intensity units 
and the color scale shows heliocentric systemic 
velocity in km s$^{-1}$. The dashed circle indicates the bridge 
region. {\it Right}: The isovelocity contours are shown from
$7009$ to $7921$ for every $20$ km s$^{-1}$. \label{fig4}
The circle indicates strong deviation from circular motions 
in the inner disk of NGC 5278.}
\end{figure*}

The total velocity field of KPG 390 with
each component of the pair and the bridge,
is shown in Fig.~\ref{fig4}
with H$\alpha$ isophotes superposed.
The outer isophote clearly shows a common 
envelope enclosing both galaxies.

The velocity field is far from regular 
and quite distorted as one can see in the 
isovelocity contours map shown in the right 
panel of Fig.~\ref{fig4}.    
In the disk of NGC 5278 the
radial velocity values are in the range
$7400-7860$ km s$^{-1}$. 
On the other hand, inside the
disk region of NGC 5279 the radial velocity values
are in the range ($7550-7650$ km s$^{-1}$).
For NGC 5279
the mean radial velocity value is of
$\approx 7600$ km s$^{-1}$.
The radial velocity values in
the bright arm region of the primary galaxy (north side of NGC 5278) 
are in the range $7350-7480$ km s$^{-1}$.
In this zone the velocity profiles are slightly broader than
those in the disk of NGC 5278.
\begin{figure}[!htp]
\epsscale{1.0}
\plotone{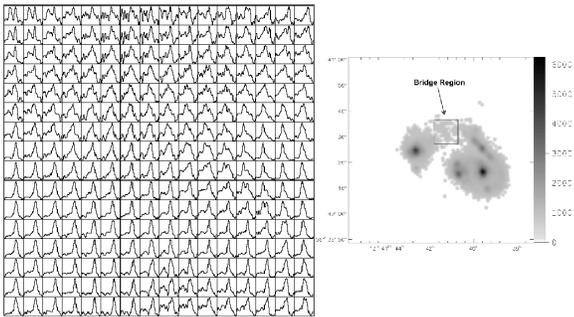} \figcaption{H$\alpha$ radial
velocity profiles showing the
region of interaction between the two galaxies,
superposed onto the H$\alpha$ image of KPG 390. 
The profiles are normalized by the corresponding intensity in 
each pixel. The square indicates the region of double profiles 
displayed on the left. These are the original spectra. \label{fig5}}
\end{figure}
\begin{figure}[!htp]
\epsscale{1.0}
\plotone{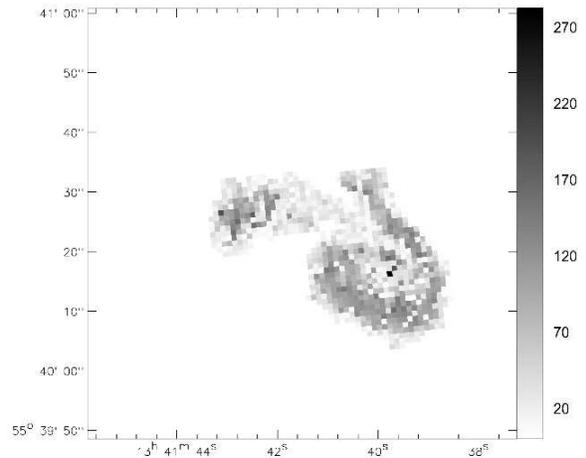} \figcaption{
Velocity dispersion field of KPG 390 obtained from the FP
velocity cubes. The color bar scale numbers are heliocentric
velocity dispersions in km s$^{-1}$.\label{fig6}}
\end{figure}

In the bridge region indicated by a dash circle in Fig.~\ref{fig4},
the radial velocity profiles are double, or distorted, and the kinematics
is more complicated. Such profiles shown in Fig.~\ref{fig5} (region inside 
the square) have a main radial velocity component and a faint secondary 
one. In this figure the north
part shows profiles with lower signal--to--noise ratio, whereas in the 
south part the profiles have higher signal to noise ratio. 
The mean radial velocity of the brightest 
peak in the double profiles of the bridge region is $\approx7600$
km s$^{-1}$, which is close to the mean systemic velocity of both galaxies. 
The faintest velocity component 
(at $7200-7330$ km s$^{-1}$) could be associated with an extended 
gas outflow due to the interaction. This outflow would be also 
responsible for the zone of peculiar velocities discussed 
in the next section.

We also present dispersion velocity map obtained from FWHM map
in Fig.~\ref{fig6}.
The map shows that wider profiles are concentrated in the 
inner disk of both galaxies and along the 
spiral arms of NGC 5278. In the bridge region the FWHM
is smaller, contrary to the broadening that is expected 
as a result of interaction.

Summarizing, we conclude that there is 
a transference of material between the two components 
of the pair, indicatating the ongoing interacion process.

\subsection{Rotation curves}

Following \citet{Fuentes-Carrera2002, Fuentes-Carrera2004}
we obtain the rotation curve of each galaxy. The rotation
curves were obtained from the corresponding velocity fields 
considering the pixels within a given angular sector along 
the major axis.
The main caution is to exclude the spurious pixels near the 
major axis leading to a strong dispersion of the points of
the rotation curve. The exclusion of those points guarantee
us the symmetry of both sides of the rotation curve.  
We can see from the velocity fields that the
inner parts of these two galaxies are not strongly
perturbed by the interaction process.
This is true at least up to a certain radius.
In the case of NGC 5278 this radius is $\approx7$
kpc ($\approx14\arcsec$) and for NGC 5279 it
is $\approx6$ kpc ($\approx12\arcsec$). 
Thus, we can accurately determine the rotation curve of both galaxies
considering a region of the velocity field within a
sector of a specified angle inside these radii.

\subsubsection{NGC 5278}\label{ngc8}
The rotation curve of NGC 5278 was obtained with pixels
in the velocity field within an angular sector 
of $20\degr$ around the galaxy major axis.  
The photometric center of this galaxy is the position of the 
brightest pixel in the continuum map.
The physical coordinates of the photometric 
center are R.A.$=13^{h} 41^{m} 39.36^{s}$ and Dec.$
=55\degr  40\arcmin 47.13\arcsec$. The kinematic center, 
derived as the position around the photometric center
at which the scatter in rotation curve is minimized, is R.A.$=
13^{h} 41^{m} 39.33^{s}$ and Dec.$=55\degr 
40\arcmin 44.39\arcsec$.
The kinematic center used to compute the rotation
curve of this galaxy matches the photometric
center within $2.4\arcsec$. We compare the 
kinematical parameters for this galaxy
with those given in the literature. 
The kinematic parameters that give us the 
most symmetric, smooth, and low-scattered 
curve inside a radius of $12\arcsec$ are 
P.A.$=(42\pm2)\degr$, {\it i} $=(42\pm 2)\degr$, 
and $V_{syst}=(7627\pm10)$ km s$^{-1}$.
Comparing the values of the kinematic parameters
with values found in the LEDA database
(see Table \ref{tbl-1}) we notice
that they differ. In this database the P.A. value is
of $53.0\degr$, the inclination with respect to
the plane of the sky is of $39.6\degr$ which is
almost the same value we find, and the
systemic velocity is $7559\pm20$ km s$^{-1}$.
These discrepancies are principally due to the different
methods used to determine these parameters
(photometric methods in the case of inclination
and position angle and slit spectroscopy methods
in the case of systemic velocity).
 
\begin{figure*}[!htp]
\epsscale{2.0}
\plotone{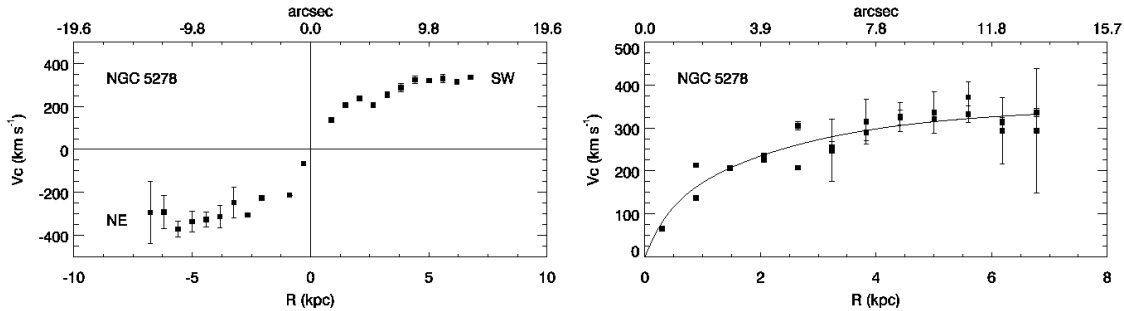} \figcaption{{\it Left:} Rotation curve 
of NGC 5278 for both 
sectors of the galaxy superposed: approaching (NE) 
and receding (SW) part. {\it Right:} 
Overplotted both parts of the rotation curve
together with an exponential fit. The error bars are dispersion 
of values within the considered sector.\label{fig7}}
\end{figure*}

\begin{figure}[!htp]
\plottwo{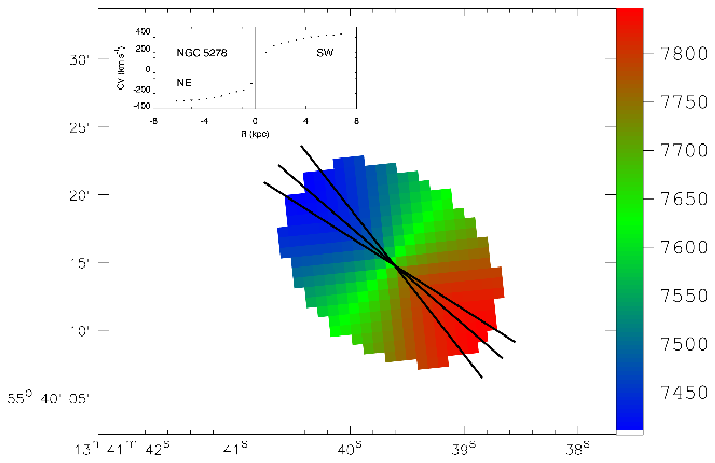}{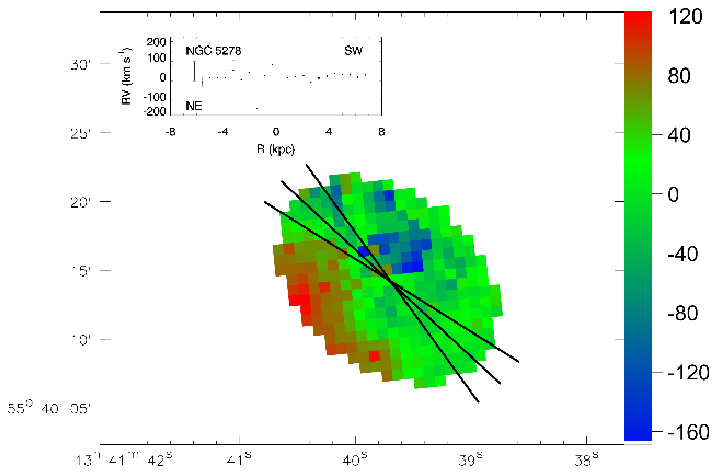} \figcaption{{\it Left}:
Synthetic circular velocity field of NGC 5278.
{\it Right}: Residual velocity field of NGC 5278.
The values are heliocentric systemic
velocities in km s$^{-1}$. The inset panels show the 
rotation curve obtained in corresponding
sectors of $20\degr$.\label{fig8}}
\end{figure}

The rotation curve shown in Fig.~\ref{fig7} 
extends only up to $6.8$ kpc (equivalent to $13.3\arcsec$),
because after this point 
the scatter is so large that the results are not 
reliable. As one can see the rotation 
curve is rather symmetric, has low scatter, and rises rapidly till 
R$=1.5$ kpc ($3\arcsec$), 
reaching the velocity of $\sim200$ km s$^{-1}$.
After this point the curve slowly reaches the maximum rotation 
velocity of $350$ km s$^{-1}$ at R$=5.8$ kpc ($11.5\arcsec$). 
The latter two points show a slow decay to $\sim300$ km s$^{-1}$. 
In order to investigate the peculiar velocities in the disk of NGC 5278,
we constructed an artificial velocity field using the fit to the 
averaged rotation curve shown in Fig.~\ref{fig7}. Then we subtracted 
the created field from the observed one, and obtain the 
residual velocity field. The resultant fields are shown in Fig.~\ref{fig8}.
The synthetic velocity field shown
in the left panel of Fig.~\ref{fig8}, was limited by the extension
of the rotation curve, and does not cover the entire velocity field of NGC 5278.
The inset shows the rotation curve obtained within the same
angular sector as the rotation curve in Fig.~\ref{fig7}.

The residual velocities shown in the right panel of 
Fig.~\ref{fig8} are quite small, except for a 
point with $\sim-167$ km s$^{-1}$ which belongs
to a strongly perturbed region located to the north of the disk 
(the inset of Fig.~\ref{fig8}). In the rotation curve fit shown 
in Fig.~\ref{fig7} this point was not included.

\begin{figure}[!htp]
\epsscale{1.0}
\plotone{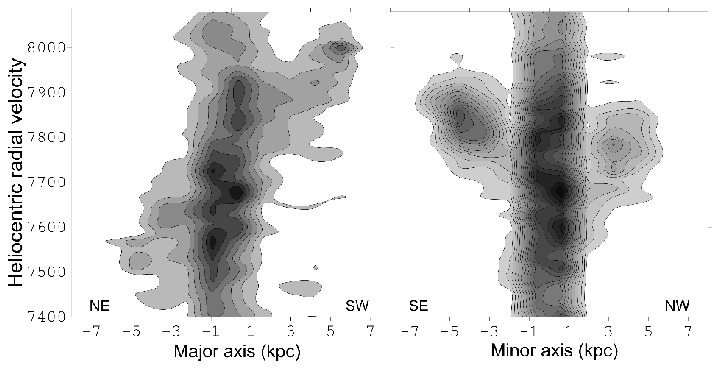}\figcaption{
Position velocity diagrams of NGC 5278 
along the kinematic major axis (left), and along 
the kinematic minor axis (right). The contours are separated by $20$ arbitrary 
units ($0.33\sigma$), and the minimum contour level 
corresponds to $1.6\sigma$.\label{fig9}}
\end{figure}

The position velocity diagrams (PVDs) extracted along 
the major and minor kinematic axes of NGC 5278 bring 
information about the symmetry of the disk.
As shown in Fig.~\ref{fig9} the PVD along the major axis
resembles the shape of the rotation curve. However, the distribution
of the H$\alpha$ intensity is not symmetric, being fainter in the NE side.
On the contrary, the PVD along the minor axis is strongly asymmetric
showing a bright region on the SE, and a fainter region on the NW.
Such asymmetries are usually associated with radial motions of 
the gas inwards or outwards. 
The bright region on the SE, corresponding to southern spiral arm, 
differs by $\approx80$ km s$^{-1}$ from the NW side.
These regions reflect the zones of large residuals observed
in Fig.~\ref{fig8}. We also extracted PVDs parallel to the minor axis at 
positions $\pm 3$ kpc from the kinematic center and found a 
similar asymmetry in velocity.

\subsubsection{NGC 5279}\label{ngc9}
The brightest pixel of NGC 5279 in our
continuum map has coordinates: R.A.$
=13^{h} 41^{m} 44.240^{s}$ and Dec.$=55\degr 
41\arcmin 1.45\arcsec$.
The coordinates of the kinematical 
center are R.A.$=13^{h} 41^{m} 43.901^{s}$ 
and Dec.$=55\degr 41\arcmin 0.32\arcsec$.
The kinematic center used to obtain 
the rotation curve matches the photometric center 
within $2.6\arcsec$.
The kinematic parameters that reduce significantly 
the asymmetry and scatter in the rotation curve were 
in this case: P.A.$=(141.5\pm1)\degr$, $i=
(39.6\pm 1)\degr$ and
$V_{syst}=(7570\pm 10)$ km s$^{-1}$.

As in the case of NGC 5278 we compare the P.A.
of the major axis, the inclination and the 
systemic velocity with the values of LEDA and  
NED database. According to LEDA the P.A.$=3\degr$ and 
the inclination is $i=59.8\degr$ 
and the NED database gives for the systemic 
velocity $7580$ km s$^{-1}$. From these data 
we see that the major discrepancies with our results are in the 
values of P.A. and inclination.
In Table \ref{tbl-1} we compare our results and the 
parameters given in the literature.

\begin{figure*}[!htp]
\epsscale{2.0}
\plotone{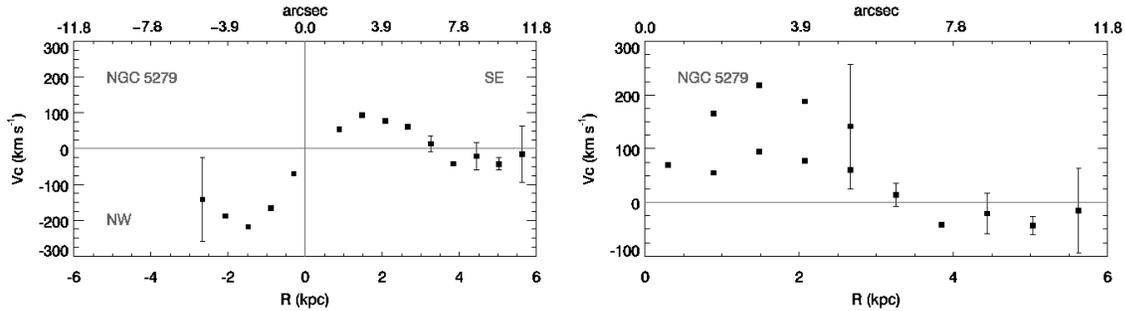} \figcaption{{\it Left:} Rotation curve 
of NGC 5279 for both 
sectors of the galaxy: approaching (NW) and 
receding (SE) side. 
{\it Right:} Overplotted both parts of the rotation curve.
The error bars represent dispersion of velocity points within 
the considered sector.\label{fig10}}
\end{figure*}

In order to minimize the asymmetry and scatter within a 
radius of $12\arcsec$ in the rotation curve 
of NGC 5279, we choose an angular sector of $20\degr$ 
around the major axis of the galaxy. 
Figure \ref{fig10} presents the rotation curve of NGC 5279: 
as an approaching (NW) and a receding (SE) side, and 
the superposition of both.
From Fig.~\ref{fig10} it is clear that the rotation curve of 
NGC 5279 has a very unusual behavior. 
Due to the lack of data points it was impossible 
to minimize the scattering and asymmetries of the rotation curve.
A high asymmetry of both parts of the rotation curve did not allow
to make a reliable fit, so we discuss them separately.

The approaching part of the rotation curve reaches the maximum velocity of
$-142$ km s$^{-1}$ at radius of $-2.7\arcsec$.
Here the velocity is of $-142$ km s$^{-1}$. It is impossible to say if 
there is an actual decrease of the rotation velocity because after the radius 
of $-5.52\arcsec$ we have no points available.
A similar behavior is observed in the receding part of the rotation curve
which reaches the velocity $\sim94$ km s$^{-1}$ at $3.3\arcsec$.
After this point the curve presents a decrease up to zero at 
the radius $5.52\arcsec$, and even shows the change of 
direction of the rotation.
Beyond this radius the determination of the rotation curve
is not possible due to high data scattering.

\begin{figure}[!htp]
\epsscale{1.0}
\plotone{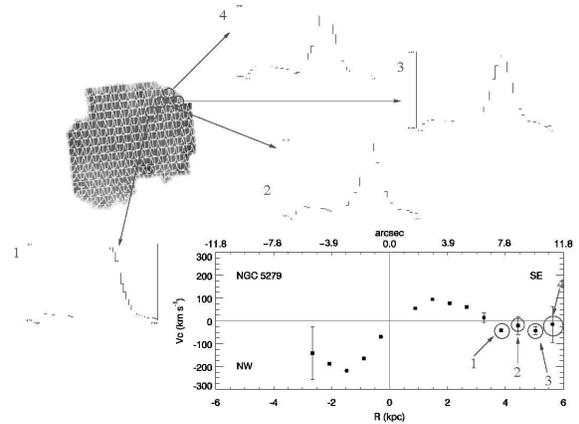} \figcaption{Velocity field 
of NGC 5279 with superposed radial velocity profiles.
Profile identification for the four negative points 
in the rotation curve.\label{fig11}}
\end{figure}

It is clear that the behavior 
of this side is very unusual.
In the receding half of the RC there are four points that are lower 
than zero. The first is at r$=7.43\arcsec$ 
and has a rotational velocity of $-42$ km s$^{-1}$. 
The second point is at r$=8.56\arcsec$ with a 
rotational velocity of $-21$ km s$^{-1}$. The third point is 
at r $=9.77\arcsec$ and its rotational 
velocity is $-43$ km s$^{-1}$. The fourth point is at
r $=11.07\arcsec$ and its rotational 
velocity is $-15$ km s$^{-1}$.
We identify the regions corresponding to these four points
in the velocity field of NGC 5279 shown in Fig.~\ref{fig11} 
and we have not found any evidence of double profiles nor 
any other anomalous issue that could help explain the 
discrepancy in the velocities of these regions.
The superposition of the H$\alpha$ image
and the velocity field of NGC 5279 helps us realize 
that these regions belong to the external part of 
the disk of this galaxy.
In this zone the material rotating around the 
NGC 5279 is more influenced by the
bridge region between the two galaxies than 
by the inner part of the disk of NGC 5279. 
A possible explanation is that material of NGC 5279 is
dragging along the bridge between the two galaxies in
opposite direction. The interaction can explain the observed high degree
of non-axisymmetry, seen in the superposition of the approaching 
and receding side of the rotation curve of NGC 5279 in 
Fig.~\ref{fig10}. 

The asymmetry of the rotation curve did not allowed us to
perform a detailed analysis of the velocity field
of NGC 5279.

\section{Mass estimates using dynamical analysis}\label{orm}
A range of possible masses was computed for each 
galaxy of this pair using an estimate described by 
\citet{Lequeux1983}. This estimation consists in 
the calculation of the mass $M\left(R\right)$
up to a certain radius where the rotation velocity
$V_c\left(R\right)$ has been measured. For spiral
galaxies, $M\left(R\right)$ is in the range 
$\kappa R V_c^2\left(R\right)/G$, 
the coefficient $\kappa\in[0.6-1.0]$ whose limit values
stand for the case of galaxies dominated by a flat disk 
or a massive spherical halo, respectively. For NGC 5278 
the maximum velocity of $350$ km s$^{-1}$ is reached 
at the radius of R$_{max}=11.5\arcsec$ 
equivalent to $5.6$ kpc. For this galaxy the mass in the 
case of a flat disk is M$^1_{flat}=9.7\times 10^{10}$ M$_{\odot}$, 
where M$^1$ indicates the mass of the primary 
galaxy. If we consider the spherical case, the mass is 
M$^1_{sph}=1.6\times 10^{11}$M$_{\odot}$.
For NGC 5279 the maximum velocity of $\sim250$ km s$^{-1}$ is 
achieved at R$_{max}=3.0\arcsec$ equivalent 
to $1.5$ kpc. For this galaxy the mass considering the 
flat disk case is M$^2_{flat}=1.3\times 10^{10}$ M$_{\odot}$ 
and in the spherical case is M$^2_{sph}=2.2\times 10^{10}$ M$_{\odot}$, 
where M$^2$ indicates the mass of the secondary galaxy. 
One should note also, that the rotation 
curve is very unusual in this case, a fact that may influence 
significantly the determination of the mass.

A second mass estimate is obtained using the relation of
\citet{Karachentsev1984} to compute the mass of the pair
from their orbital motion, assuming a circular orbit. 
\begin{equation}\label{eq1} 
M_{orbital}=\frac{32}{3 \pi} \left(\frac{
\Delta V^2 \times X_{12}}{G}\right),
\end{equation}
where $\Delta V$ is the difference between the 
systemic velocities of the galaxies, $X_{12}$ is the 
projected separation between the nucleus of each galaxy,
$G$ is the gravitational constant, and $32/3\pi$ is the 
mean value of the projection factor for circular motion 
of the members of the pair and isotropic orientation of 
the orbits. For KPG 390 we have $\Delta V=57$ km s$^{-1}$ and 
$X_{12}=16.6$ kpc. Assuming that $\Delta V$ corresponds to
a lower limit value for the orbital motion 
of the pair, we obtain $M_{orbital} \ge 4.27 \times 
10^{10}$ M$_{\odot}$. This is less than the values obtained for 
M$^1$, following the Lequeux method. Moreover, the sum of the individual 
masses of galaxies in the spherical case is $1.8\times 10^{11}$ 
M$_{\odot}$ while in the flat disk case is $1.1 
\times 10^{11}$ M$_{\odot}$. In both cases M$_{orbital} 
<$M$_{sum}$ and the ratios between these values is 
M$_{sum}/$M$_{orbital}=4.3$ in the spherical case and 
M$_{sum}/$M$_{orbital}=2.6$ in the disk flat case. Consequently,
the orbital mass method based on statistical grounds
does not allow an accurate estimation of the masses.
On the other hand, the Lequeux estimation is limited 
by the radial range adopted in the calculation of the 
galaxy mass. Thus, none of the above approaches allows us
a precise mass estimation.

\section{Mass computation from rotation curve 
decomposition of NGC 5278}\label{mcrcd}
In order to accomplish the rotation curve decomposition of NGC 5278, 
we consider that the galaxy has two components that contribute to 
the rotation curve: an exponential disk and a massive dark matter 
(DM) halo. The disk  was assumed to be thin and not truncated with 
an exponential density distribution \citep{Freeman1970} expressed 
by the relation:
\begin{equation}
\label{eq2}
\Sigma=\Sigma_0 e^{-r/h},
\end{equation}
where $\Sigma$ is the surface density of the disk,
$\Sigma_0$ is the central surface density and $h$ is the
scale length of the disk. 

Photometric observations provide the surface brightness profiles
from which we can obtain the central surface brightness, 
$\mu_0$, in magnitude units and the disk scale lenght in kpc.
In order to transform these observable parameters to mass 
density distribution, it is assumed that the $M/L$ ratio is uniform and
constant over the disk. In principle, the disk $M/L$
could be known from photometric and spectroscopic observations 
of the disk which allow us to know the colors, or to perform a 
population synthesis analysis.

We tested three different types of DM halos: Hernquist halo \citep{Hernquist1990},
Navarro, Frenk $\&$ White halo (NFW) \citep{Navarro1996}, and
spherical pseudo-isothermal halo. The density profile of the Hernquist halo is given by:

\begin{equation}\label{eq3}
\rho\left(r\right)=\frac{M}{2\pi}
\frac{a}{r}\frac{1}{\left(r+a\right)^3}
\end{equation}

\noindent where $M$ is the total mass and $a$ is a
length scale \citep{Hernquist1990}. The density profile of the NFW halo is
\citep{Navarro1996}:
\begin{equation}\label{eq4}
\rho\left(r\right)=
\frac{\delta_c\rho_{crit}}{\left(r/r_s\right)\left(1+r/r_s\right)^2},
\end{equation}
where $r_s=r_{200}/c$ is a virial radius and $\rho_{crit}=3 H_0^2/8\pi G$
is the critical density, the Hubble constant $H_0$=72 km s$^{-1}$ Mpc$^{-1}$,
$\delta_c$ and $c$ are dimensionless parameters 
which are mutually related.
The mass of the halo is determined by $r_{200}$ with the relation $M_{200}=200\rho_{crit}
\left(4\pi/3 r_{200}^3\right)$.
The spherical pseudo-isothermal halo has a density profile given by the relationship:
\begin{equation}\label{eq6}
\rho\left(r\right)=\rho_0\left[1+\left(
\frac{r}{r_c}\right)^2\right]^{-1},
\end{equation}
where $\rho_0$ is the central density of the halo, and
$r_c$ the core radius of the halo. 

To accomplish the rotation curve decomposition
we fit the rough rotation curve data using 
an exponential function (Fig.~\ref{fig7}).
Then we use this fit to perform the
disk-halo decomposition. We perform the rotation curve 
decomposition only for NGC 5278 using the photometric data given by 
\citet{Mazzarella1993} and the 
LEDA\footnote{http://leda.univ-lyon1.fr} database \citep{Paturel2003}.
\citet{Mazzarella1993} present the values of the Holmberg radius
of the stellar disk of the galaxy, the mean B-band surface
brightness inside this radius, and the apparent magnitude in the B-band
for both galaxies.
The LEDA database provides the length of the projected major axis
of the galaxy at the isophotal level $25$, and the corresponding B-band
surface brightness. From these two relations we obtain the central surface brightness, 
$\mu_0=20.5$ mag arcsec$^{-2}$ and disk scale length of $11$ kpc. 
Note, however, that the latter value is greater than the extension 
of the rotation curve shown in Fig.~\ref{fig7}, 
so we treat it as an upper limit.
Since for this galaxy we do not have any robust restiction on
the luminous mass distribution, we vary the disk scale radius 
$h\in[1.0-11.0]$ kpc and the mass to luminosity ratio $M/L_B\in[1.3-6.3]$.
The halo mass and the halo scale length are in the range 
$[0-10^{13}]$ M$_{\odot}$ and $[0-20]$ kpc, respectively.

In order to carry out the rotation
curve fit we use the IDL MPFIT\footnote{http://purl.com/net/mpfit}
package \citep{Markwardt2009}\footnote{The original fortran package that implements
Levemberg-Marquardt method
was developed by J. Mor\'e and collaborators at Argonne National
Laboratories \citep{More1978}}.
As an input to the routine we pass the fit to the rotation curve 
of NGC 5278 and a weights vector,
computed as an inverse of squared standard deviations. The outputs
are the best fit model and the $\chi^2$ normalized by the number of 
degrees of freedom and the corresponding weights.

\begin{figure}[!htp]
\epsscale{1.0}
\plotone{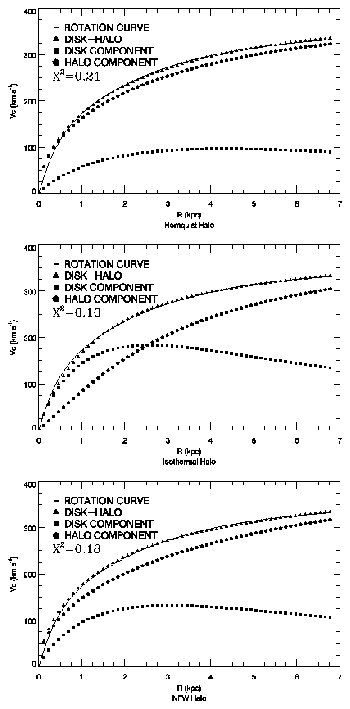}\figcaption{Disk-Halo 
decomposition of the rotation curve of
NGC 5278.\label{fig12}}
\end{figure}

The rotation curve of NGC 5278 can be well fitted with
spherical pseudo-isothermal halo,
Hernquist halo and NFW halo (Fig.~\ref{fig12}).
The results obtained are summarized in Table \ref{tbl-3}.

For pseudo-isothermal halo the best fit was obtained with a 
mass to light ratio of $2.8$ and the disk mass $2.42\times10^{10}$ M$_{\odot}$. 
In the case of the Hernquist halo we choose $M/L=1.3$, which corresponds to
the disk mass $1.13\times10^{10}$M$_{\odot}$.
For the \citet{Navarro1996} halo (NFW halo) the $M/L$
ratio was of $1.7$ and the disk mass of 
$1.5\times10^{10}$ M$_{\odot}$. 

The pseudo-isothermal mass is in agreement with the dynamical mass 
estimate of Section \ref{orm} and may be considered as a lower limit
to the mass of NGC 5278. The second and the third value
do not agree with the first estimation of masses 
and can be considered as upper limits to the mass of NGC 5278.
The three fits roughly reproduce the observed rotation 
curve while the pseudo-isothermal halo requires a dark matter amount 
ten times smaller than the other two halo models.
In all three models the halo component dominates the mass of NGC 5278. 

In order to corroborate our results we also estimate the dynamical mass of
NGC 5278 using the Tully-Fisher relation \citep{Tully1977}. This relation 
can be written in the following form M$_{dyn}=V^4_{max}/2GcH_0$, where
$V_{max}=350$ km s$^{-1}$ is the maximum circular velocity, G is the
gravitational constant, $c$ is the speed of light and $H_0$ is the Hubble
constant at current time. We found a dynamical mass of 
$8.1\times10^{10}$ M$_{\odot}$. This value agrees with the dynamical
mass estimate in the case of a flat disk (Section \ref{orm}).  

In the course of the fitting
process, we have also explored the effects of having as a
third component a buldge with the Hernquist
density profile. However, the bulge mass and the length
scale in all three cases considered below were too small ($3.2\times10^8 M_{\odot}$
and $0.6$ kpc for spherical isothermal halo and $1.8\times10^8$
$M_{\odot}$ and $0.3$ kpc for Hernquist halo and NFW halo)
to justify the necessity of a bulge component in total mass estimation.

The total mass obtained only with our kinematical data does not ensure
the actual mass distribution of NGC 5278, because of the small range of H$\alpha$ 
rotation curve, which reflects only the dynamic of the inner disk.    
Photometric, spectroscopic and population synthesis studies are 
required to fix the disk $M/L$ ratio and thus discriminate between 
the different density distributions of DM halos as well as the 
ratio between luminous and dark matter in this galaxy.

Concerning NGC 5279, we were unable to make a reasonable fit to the rotation
curve, and consequently, estimate its mass, due to its unusual shape and
lack of resolution.

\begin{table*}[ht]
\begin{minipage}[t]{\columnwidth}
\caption{Mass determination from rotation curve decomposition.}
\label{tbl-3}
\begin{center}
\renewcommand{\footnoterule}{}  
\begin{tabular}{p{5.0cm} p{3.5cm} p{3.5cm} p{3.5cm}}
\tableline
\tableline
Rotation curve mass & Pseudo-isothermal\tablenotemark{a} & Hernquist& NFW\\
Disk component (M$_{\odot}$) & 2.42$\times$10$^{10}$ &
1.13$\times$10$^{10}$ 
& 1.5$\times$10$^{10}$ \\
Disk lenght scale (kpc) & 1.2 
& 2.0  & 1.42\\
Disk M/L & 2.8  & 1.3 & 1.6\\
Halo component (M$_{\odot}$)& 1.9$\times$10$^{11}$ &
2.1$\times$10$^{12}$ 
& 6.3$\times$10$^{12}$\\
Halo lenght scale (kpc) & 2.8 
& 17.7  & 16.8\\
Total Mass (M$_{\odot}$)& 2.1$\times$10$^{11}$  &
2.1$\times$10$^{12}$ & 6.3$\times$10$^{12}$ \\
$\chi^2$\tablenotemark{b} & 0.12 & 0.53 & 0.54\\
\tableline
\tableline
\end{tabular}
\tablenotetext{a}{Maximum Disk}
\tablenotetext{b}{Normalized $\chi^2$ by $58$ degrees of freedom.}
\end{center}
\end{minipage}
\end{table*}

\section{Discussion}\label{dsc}
To study in detail the kinematic structure of a particular isolated 
pair of galaxies it is very important to have two 
dimensional velocity fields for each pair component. 
The kinematic and dynamical analysis allows us to 
obtain useful information about the transference 
of material between the two galaxies as well as about 
the principal mass components of the system.

The procedure of disk-halo decomposition
of a given rotation curve is based on a criterion used to decide how 
much mass has to be assigned to the luminous components and with what 
distribution. In the case of Sc galaxies that do not involve 
a significant bulge component, this problem reduces to fit 
the observed rotation curve only with two contributions: a 
disk component and a halo component. The profile of the 
contribution of the rotation curve due to the disk is, in principle, fixed 
in shape by the observed photometry, while its scale is
controlled by the mass--to--light ratio assigned to the disk.     
The difference between the observed rotation curve and the 
disk contribution is attributed to the presence of unseen
matter i.e. a dark halo component. There are distinct solutions
to the problem of modelling the mass distribution of spiral
galaxies. A conservative effort is to minimize the role of unseen
matter that leads to the choice of the maximum-disk solution for the
disk-halo decomposition \citep{van$-$Albada1985}. From a 
dynamical point of view, other solutions, with a smaller value
for the disk mass--to--light ratio and a heavier dark halo are 
also viable. What regards the luminous part of the rotation curve of  
NGC 5278, this galaxy is classified as an Sb galaxy, thus, in 
principle it could have a significant bulge contribution, 
but with our analysis we demonstrate that the bulge component
is too small to need to be taken into account in our decomposition. 
On the other hand, the disk 
contribution is partially set by the available photometric data
and in the maximum-disk case (pseudo-isothermal halo) is of
fundamental importance for the mass distribution.  
As what concerns the dark component of NGC 5278 we can notice that
the difference in the halo mass between pseudo-isothermal and
Hernquist or NFW halo components is a factor of ten in the mass. 
This could be partially due to the choice of the maximum-disk component
in the case of pseudo-isothermal halo.
The Hernquist and NFW halo
components have a very similar behaviour as expected.
From our RC fits, there is no reason to think that one of the three
fits represents better than the others the mass distribution 
of NGC 5278. The rotation curve decomposition 
performed here is only a first attempt at modelling the
mass distribution of NGC 5278 that should be completed with at 
least a detailed photometry of this galaxy.

On the other hand, this kinematic study sheds light on the 
geometry of the galaxy encounter by determining univocally 
the real orientation in the sky of the galaxy members, as well 
as the kind of spiral arms they possess. This later point is 
not irrelevant in the case of interacting systems where a 
possibility of having leading spiral arms is open.
Indeed, even if leading spiral arms in galaxies are a very uncommon
phenomenon, the only examples where are found are 
interacting systems. Let us revise problem of leading 
or trailing arms. \citet{deVaucouleurs1958}, in a seminal paper, 
gave a series of criteria in order to determine the true 
orientation of a galaxy in space and the kind of spiral 
arms it has. He also established that in a sample
of spiral galaxies the whole trial had trailing arms, 
concluding that the spiral arms are all trailing, at least 
for galaxies considered isolated. However, this last issue 
has been questioned and some authors have revised this 
important conclusion of the de Vaucouleurs work. The first
systematic study on the topic was done by \citet{Pasha1982} and
\citet{Pasha1985}. These authors from a sample of 200 spiral
galaxies found 4 leading arm galaxy candidates. 
\citet{Sharp1985} have shown that one of this four leading 
arm candidates (in NGC 5395) is not leading. The other
three candidates are highly dubious cases regarding orientation and
existence of arms (NGC 4490) and their tilt (NGC 3786 and NGC 5426).
\citet{Byrd2008} demonstrate that NGC 4622 has leading arms since
two pairs of detected arms have the opposite sense with each other.
\citet{Grouchy2008} detected another example of leading spiral
structure in ESO 297-27. Thus, as one can see, there are not so many
cases of leading spiral arms in the literature.
Following \citet{Sharp1985} there is a criterion 
that determines if any particular spiral galaxy has trailing 
or leading arms. This criterion is based on three main 
clues (receding-approaching side, direction of spiral arms
and the tilt of the galaxy, i. e., which side is closer to observer). 

\begin{figure}[!htp]
\plotone{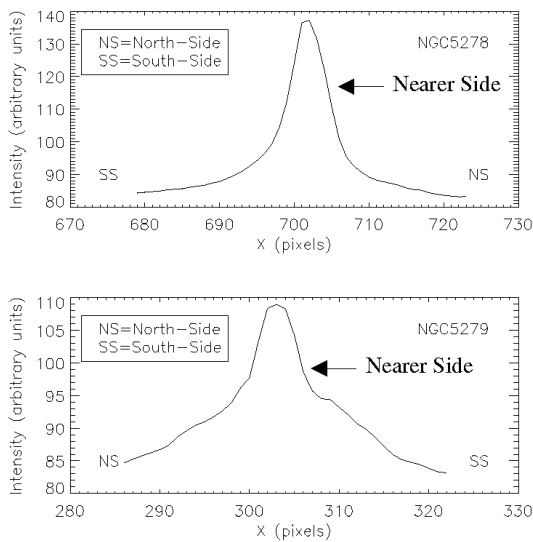} 
\epsscale{1.2}
\figcaption{Intensity profiles along the 
kinematic minor axis of
NGC 5278 and NGC 5279. The cross--section is captured
in the Hubble image (Fig.~\ref{fig2}). From the top
profile it is clear that the northern--side is the nearer
in NGC 5278, because the profiles fall more abruptly
than along the southern--side. In the case of NGC 5279
the southern--side is the nearer because the profiles
fall more abruptly than along the northern--side, as
one can see from the bottom profile.\label{fig13}}
\end{figure}

\begin{figure}[!htp]
\epsscale{0.8}
\plotone{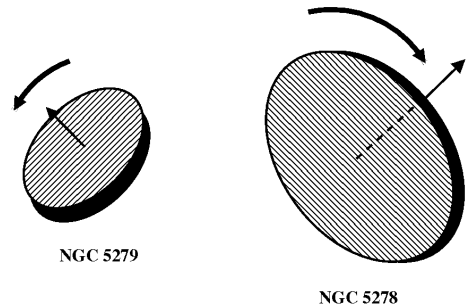} 
\figcaption{A possible spatial configuration of KPG 390.\label{fig14}}
\end{figure}

In our particular case, we have both the kinematic information 
in order to establish which side of the galaxy is receding and 
which side approaching, as well as very conspicuous morphological 
aspects such as well defined spiral arm patterns and the presence 
of dust lanes in both galaxy members running near the galaxy nuclei. 
This last issue will be used in what follows in two main ways: 1) 
we will suspect that the nearest side of the galaxy is the side 
hosting the dust lane and, 2) we will check it by getting an 
intensity profile of the galactic nucleus along the minor axis. 
In this kind of profiles, the nearest side is the steepest one 
(because of the presence of the dust lane).
In the case of NGC 5278,
the receding radial velocities are in the south-western part, while the 
approaching radial velocities are at the north-eastern side. 
From Fig.~\ref{fig2} (bottom panel) it is clear that the arms of
NGC 5278 point in anti-clockwise direction and the dust lane is 
located at the concave side of the bulge, thus the northern side is 
the nearest. This fact is confirmed by the profile extracted
along the kinematic minor axis of NGC 5278 
(see Fig.~\ref{fig13}). 
From these figures and the above criteria we have decided that NGC 5278 is
a trailing spiral because the sense of rotation is opposite to 
the direction of the arms. We were able to apply similar 
arguments to NGC 5279. In this case the receding radial 
velocities are at the north-eastern side 
of the galaxy and the approaching radial velocities are at 
the south-western part. The arms of NGC 5279 point in clockwise direction and 
the nearest side is the southern side. As in the case of 
NGC 5278 this fact is confirmed by the profile extracted
along the kinematic minor axis of NGC 5279 (see Fig.~\ref{fig13}) 
\citep{Vaisanen2008}. We can conclude that
NGC 5279 is a trailing spiral also because the sense of rotation 
is opposite to the direction of the arms. 
A scheme of 3D orientation of KPG 390 derived from our 
kinematic analysis is shown in Fig.~\ref{fig14}. 
We estimate a lower limit to the time scale of interaction 
using the ratio of the projected separation between both 
galaxies ($16.6$ kpc) \citep{Karachentsev1972} and the 
difference in the systemic velocities of the components of 
the pair ($57$ km s$^{-1}$). The result was of $2.9\times 10^8$ yrs. 
This fact agrees very well with the morphology of the pair. 
In fact, NGC 5279 do not appear very distorted by the 
interaction and NGC 5278 shows some signs of an undergoing 
interaction process. 
This galaxy presents an extended spiral arm toward the 
companion galaxy and between the two galaxies one can 
see a bridge-like structure probably generated by 
the interaction process. From the age of interaction 
we can say that this is an early-stage interaction.

We do not know if the bar in the secondary galaxy was
due to the interaction process. The origin of 
bars in interacting disk galaxies remains unclear. 
Though there have been works showing statistical 
evidence that companions trigger the formation of 
bars \citep{Elmegreen1990, Marquez2000}, bars can 
be formed as a result of disk instabilities 
regardless of their environment \citep{vandenBergh2002}. 
Perhaps this bar existed before the 
interaction process took place. Realistic numerical 
simulations taking into account the kinematic restrictions 
are needed to establish the extent to which the interacting 
process is related to the formation of a possible 
bar in NGC 5279.

Another interesting feature is the bridge-like structure 
that is very dim in our observations. 
The fact that the bridge region emits weak diffuse
line emission enforce the idea that this structure 
is a consequence of the interaction process and 
that the transfer of material is effective in 
this zone. This area shows double velocity profiles, 
with velocities in the range of $7200-7300$ km s$^{-1}$. 
Perhaps the transfer of material between the two galaxies of the pair
might generate such double profiles. On the 
other hand, strong  motions due
to the nearness of the arm of NGC 5278 might
alter the velocity profiles in the bridge area
and originate double and broader profiles, due to a second 
velocity component related to the transfer of gas from 
one galaxy to the other.
As pointed out in Section \ref{velf}, the 
radial velocity values of the double peaks
in the bridge region are different from 
the radial velocity values belonging to
the disks of both components of KPG 390.

Regarding non-circular motions, the residual 
velocity map for NGC 5278 reveal large values in the 
northern region of this galaxy.
The large absolute values of the residual velocities 
seen in the north-western region and also along the 
inner part of the south-eastern arm 
could indicate the response of the gas 
to the passage of the companion. This can also be noticed
from the PVDs extracted along the minor axis.
In this sense, the non-typical rotation curve of 
the companion, NGC 5279, could be a result of interaction.
However, such 
a difference could also be associated with an intrinsic 
asymmetry of the disk such as a warp.
We neither can exclude the possibility of an artifact in 
the determination of the rotation curve,
given the small number of points and the large dispersion
associated to this curve. 

The results of this work is worth to compare with 
a widely studied interacting grand-design system M51.
\citet{Salo2000} with N-body simulations were able 
to match the observed morphological and kinematical data and 
derive the orbital parameters and the interaction scenario.
They explained the formation of the spiral structure of the main galaxy,
tidal tail and bridge as a result of multiple encounters.
In a previous study of a similar M51-type system
ARP 86 by the same authors, it is shown that the material transfer 
between the components could explain the large activity of the companion \citep{Salo1993}.
They conclude that such interactions produce an open spiral structure of the primary
galaxy, and the bridge tends to be directed to the companion.
The multiple passage scenario discussed in \citet{Salo2000}
produces significant peculiar velocities out of the plane of the disk,
leading to disturbances in the rotation curve, similar to 
those observed for NGC 5279. According to separate studies aimed at the
investigation of evolution of such asymmetries in rotation curves 
\citep[][and references therein]{Pedrosa2008}, 
the perturbations are short--lived ($\lesssim 1$ Gyr), suggesting 
that the encounter for KPG 390 took place within this period.

It is very difficult to obtain a clear picture of the 
process of interaction in this pair of galaxies only 
from the kinematical analysis. We require more information 
about the stellar component by means of photometry, spectroscopy and 
population synthesis models. We also need a detailed 
dynamical analysis, that would 
encompass the knowledge of the mass components 
of both galaxies of KPG 390 and  numerical 
simulations of the pair. In particular, numerical 
simulations will allow us to have the time evolution 
of the interaction process. By comparing the results 
of numerical simulations with observations 
we will gain some insight about the present stage of the 
interaction process. Surely, this study does 
not dissipate all doubts about this pair but 
may be considered like a valid approach to better 
understand this kind of phenomena.     

\section{Conclusions}\label{cls}
In this article we presented Fabry-Perot observations 
of the isolated pair of galaxies NGC 5278/79 
(Arp 239, KPG 390) showing that for an interacting 
and asymmetric system it is important 
to have kinematic information of large portions of 
the galaxies participating in the interaction process. 
We calculate the mass
of each galaxy of the pair, following several methods,
and also the lower limit to the orbital mass of the pair.
We perform the decomposition of the rotation curve for
NGC 5278 and determine the content of dark matter
of this galaxy, using different types of halos
(pseudo-isothermal, Hernquist and NFW halo).
According to our estimations the minimum mass ratio 
of NGC 5278 to NGC 5279 is $\approx7$.

We obtain the rotation curves for NGC 5278 and NGC 5279. 
From the analysis of the velocity field, 
we found the presence of double profiles
in the bridge region and in the outer regions
of both galaxies. We conclude that the 
presence of such features is undoubtedly connected 
with the interaction process.
Though this seems to be a relatively early encounter, 
several morphological features of each galaxy were 
associated with the interaction process, such as  
the $m$=1 mode of NGC 5278, the presence of bar in NGC 5279 
and the bridge connecting the two components of the pair. 
We will use the kinematic information as a starting point 
in future numerical simulations of this pair.

\acknowledgments R. P. acknowledges CONACYT for
doctoral scolarship. R. G. acknowledges CONACYT for
postdoctoral scolarship. 
This work was also supported by DGAPA-UNAM grant: IN102309 
and CONACYT grant: 40095-F.
This research has made use of 
the NASA/IPAC Extragalactic Database (NED) which 
is operated by the Jet Propulsion Laboratory, 
California Institute of Technology, under contract 
with the National Aeronautics and Space Administration.
We acknowledge the usage of the 
HyperLeda database (http://leda.univ-lyon1.fr).

\nocite{*}
\end{document}